\documentclass[sn-mathphys-num]{sn-jnl}% Math and Physical Sciences Numbered Reference Style 
\usepackage{graphicx}%
\usepackage{multirow}%
\usepackage{amsmath,amssymb,amsfonts}%
\usepackage{amsthm}%
\usepackage{mathrsfs}%
\usepackage[title]{appendix}%
\usepackage{xcolor}%
\usepackage{textcomp}%
\usepackage{manyfoot}%
\usepackage{booktabs}%

\usepackage{listings}%

\usepackage{amsmath,amsfonts,amsbsy,amssymb,mathtools,amsthm,tabulary,graphicx,times,fancyhdr, mathrsfs, pgf, epsfig, epstopdf, lscape, subfig,booktabs, enumitem,lipsum,tkz-graph,scalerel,stackengine,setspace,kbordermatrix,adjustbox,float,caption}

\theoremstyle{thmstyleone}%
\theoremstyle{thmstyletwo}%

\theoremstyle{thmstylethree}%

\begin{document}

\title[Modeling Regime Structure and Informational Drivers of Stock Market Volatility via the Financial Chaos Index]{Modeling Regime Structure and Informational Drivers of Stock Market Volatility via the Financial Chaos Index}


\author*[1]{\fnm{Masoud} \sur{Ataei}}\email{masoud.ataei@utoronto.ca}

\affil[1]{\normalsize\orgdiv{Department of Mathematical and Computational Sciences},\\ \orgname{University of Toronto}, \state{Ontario}, \country{Canada}}

%\affil[3]{\orgdiv{Department}, \orgname{Organization}, \orgaddress{\street{Street}, \city{City}, \postcode{610101}, \state{State}, \country{Country}}}


\abstract{
This paper investigates the structural dynamics of stock market volatility through the Financial Chaos Index, a tensor- and eigenvalue-based measure designed to capture realized volatility via mutual fluctuations among asset prices. Motivated by empirical evidence of regime-dependent volatility behavior and perceptual time dilation during financial crises, we develop a regime-switching framework based on the Modified Lognormal Power-Law distribution. Analysis of the FCIX from January 1990 to December 2023 identifies three distinct market regimes, low-chaos, intermediate-chaos, and high-chaos, each characterized by differing levels of systemic stress, statistical dispersion and persistence characteristics. Building upon the segmented regime structure, we further examine the informational forces that shape forward-looking market expectations. Using sentiment-based predictors derived from the Equity Market Volatility tracker, we employ an elastic net regression model to forecast implied volatility, as proxied by the VIX index. Our findings indicate that shifts in macroeconomic, financial, policy, and geopolitical uncertainty exhibit strong predictive power for volatility dynamics across regimes. Together, these results offer a unified empirical perspective on how systemic uncertainty governs both the realized evolution of financial markets and the anticipatory behavior embedded in implied volatility measures.

 }




\maketitle

\newpage
\section{Introduction}

Financial markets evolve through complex dynamics characterized by intermittent episodes of elevated volatility, systemic fragility, and abrupt regime shifts. While traditional asset pricing models often assume stationary or smoothly varying stochastic processes, empirical observations indicate that volatility undergoes sudden transitions triggered by macroeconomic dislocations, policy shifts, or geopolitical shocks. Capturing the statistical architecture of these transitions, and understanding the informational forces that drive them, remains a central challenge in the mathematical modeling of financial systems.

In this study, we investigate the temporal evolution of market volatility through the lens of the Financial Chaos Index (FCIX), a tensor- and eigenvalue-based construct designed to quantify realized volatility by measuring mutual fluctuations among asset prices~\citep{ataei2021theory}. Unlike traditional scalar measures of volatility, the FCIX captures higher-order interdependencies across asset returns, offering a richer geometric representation of systemic risk in financial networks. A longitudinal analysis of the FCIX from January 1990 to December 2023 reveals pronounced nonstationarity, characterized by phases of relative stability punctuated by sharp surges in volatility aligned with major financial crises and systemic disruptions.

Given the evident temporal heterogeneity of FCIX realizations, it becomes crucial to identify distinct regimes within the volatility process. To formally characterize structural shifts in market behavior, we employ retrospective change-point detection methods grounded in nonparametric, kernel-based frameworks. These approaches identify statistically homogeneous intervals without requiring strong parametric assumptions about the underlying stochastic process. Existing methods for regime detection, such as Markov switching models~\citep{hamilton1989new} or hidden Markov volatility models~\citep{ryden1998stylized}, often impose restrictive structure on the data-generating process. By contrast, the segmentation strategy used in this paper allows for flexible, model-free detection of regime boundaries, thereby accommodating the complex, heavy-tailed, and high-dimensional nature of financial volatility dynamics; see~\cite{truong2020selective} for further methodological details.

Financial crises are often accompanied by a perceptual dilation of time within markets, wherein information dissemination, liquidity adjustments, and volatility responses accelerate relative to normal conditions. From the perspective of market participants, short-term events acquire amplified significance and exert disproportionate influence over price dynamics. Mathematically, this phenomenon suggests that the effective market clock evolves stochastically, with random interruptions or accelerations during periods of systemic stress.  To capture this behavior, we posit that volatility evolves as a multiplicative process subject to random stopping times, leading naturally to the {Modified Lognormal Power-Law} (MLP) distribution \citep{basu2004power,basu2015mlp}, originally introduced in the context of modeling stellar and substellar mass distributions. The MLP framework integrates a lognormal core, reflecting the compounding of small shocks during stable periods, with a power-law tail that captures the heavy-tailed behavior induced by crises. This dual structure provides a principled statistical model for the coexistence of routine fluctuations and critical transitions in market behavior.

Empirical investigations have further established a strong positive correlation between FCIX and the VIX index, the latter serving as a canonical measure of implied equity market volatility~\citep{ataei2021theory}. Causality analysis indicates a bidirectional relationship between FCIX and VIX: realized volatility fluctuations captured by FCIX predict subsequent changes in VIX, while anticipatory shifts in VIX forecast future realized market volatility. This entanglement underscores that FCIX encapsulates not merely historical asset fluctuations but also elements of forward-looking risk perception embedded in market behavior. Consequently, modeling the dynamics of FCIX provides insights into both the retrospective evolution of volatility and the contemporaneous formation of systemic expectations.

Building upon the segmented regime structure of the FCIX, we further examine the informational forces that drive fluctuations in implied market volatility. It is well-established that volatility responds sensitively to the flow of economic news, policy announcements, and investor sentiment, with unexpected macroeconomic developments frequently triggering volatility spikes, particularly during periods of systemic uncertainty~\citep{baker2016measuring,al2019economic,audrino2020impact}. Recent advances have also employed machine learning techniques to forecast volatility using macroeconomic variables, sentiment indicators, and measures of economic uncertainty~\citep{bianchi2021bond}. 

Motivated by these findings, we employ an elastic net regression model to predict forward-looking volatility, as proxied by the VIX index, using sentiment-based predictors derived from the Equity Market Volatility (EMV) tracker~\citep{baker2019policy}. By linking thematic categories of economic, financial, policy, and geopolitical uncertainty to volatility dynamics across the FCIX-inferred regimes, we trace how shifts in market expectations correspond to transitions between structural states of disorder. In doing so, we seek to identify not only the dominant sources of informational instability under varying market conditions, but also the mechanisms through which systemic stress propagates and amplifies. This predictive framework complements the regime-based analysis of FCIX by elucidating the specific uncertainty channels that underpin regime shifts, offering a unified perspective on both the endogenous evolution and exogenous modulation of financial market turbulence.

The remainder of this paper is organized as follows. Section~\ref{Sec:FCIX} introduces the construction of the FCIX and presents its segmentation from 1990 to 2023 into statistically homogeneous regimes. Section~\ref{Sec:Regime} develops the MLP framework to model regime-dependent volatility behavior and characterize the stochastic structure of financial turbulence. Section~\ref{Sec:Forces} investigates the informational drivers of implied volatility, employing sentiment-based predictors within an elastic net regression framework to trace the role of macroeconomic, policy, and geopolitical uncertainty across regimes. Finally, Section~\ref{Sec:Conclusion} summarizes the main findings and discusses potential extensions and applications of the proposed modeling approach for financial monitoring and systemic risk assessment.

\subsection{Preliminaries}

Throughout the paper, scalar quantities, vectors, and matrices are denoted using lowercase lightface (e.g., $x$), lowercase boldface (e.g., $\mathbf{x}$), and uppercase boldface (e.g., $\mathbf{X}$) letters, respectively, where all vectors are presumed to be column vectors unless otherwise stated. Boldface calligraphic letters (e.g., $\boldsymbol{\mathcal{X}}$) are used to denote tensors, i.e., multidimensional arrays generalizing matrices to higher-order structures. We use $\mathbb{R}_{\geq 0}$ and $\mathbb{R}_{>0}$ to denote the sets of nonnegative and strictly positive real numbers, respectively.

A tensor of order $N$ is an element of the space $\mathbb{R}^{I_1 \times I_2 \times \cdots \times I_N}$, where $I_n$ denotes the dimensionality along mode $n$, and the total number of elements is $K = \prod_{n=1}^{N} I_n$. Analogous to matrix rows and columns, a {mode-$n$ fiber} of a tensor is obtained by fixing all indices except the $n$th, whereas fixing all but two indices yields a two-dimensional slice, similar to a matrix cross-section. The Frobenius norm of a tensor $\boldsymbol{\mathcal{X}}$ is defined as follows:
\begin{equation}
	\|\boldsymbol{\mathcal{X}}\|_F = \sqrt{\langle\boldsymbol{\mathcal{X}}, \boldsymbol{\mathcal{X}}\rangle} = \sqrt{ \sum_{i_1=1}^{I_1} \sum_{i_2=1}^{I_2} \cdots \sum_{i_N=1}^{I_N} x_{i_1 i_2 \ldots i_N}^2 }.
\end{equation}

An important technique in tensor analysis is the {polyadic decomposition}, which approximates a tensor by a sum of rank-one tensors, also referred to as atoms. Specifically, given a positive integer $R$, the approximation is expressed as follows:
\begin{equation}
	\label{Eq:CPD}
	\boldsymbol{\mathcal{X}} \approx \widehat{\boldsymbol{\mathcal{X}}} = \sum_{r=1}^{R} \mathbf{a}_r \circ \mathbf{b}_r \circ \mathbf{c}_r,
\end{equation}
where the symbol "$\circ$" denotes the vector outer product. If $R$ corresponds to the minimal number of terms required to exactly reconstruct the tensor, the decomposition \eqref{Eq:CPD} is referred to as the {canonical polyadic decomposition}.

A tensor $\boldsymbol{\mathcal{X}}$ is said to be of rank one if it can be written as the outer product of three vectors; i.e.,
\begin{equation}
	\boldsymbol{\mathcal{X}} = \mathbf{a} \circ \mathbf{b} \circ \mathbf{c},
\end{equation}
where $\mathbf{a} \in \mathbb{R}^{I_1}$, $\mathbf{b} \in \mathbb{R}^{I_2}$, and $\mathbf{c} \in \mathbb{R}^{I_3}$.

The tensor-based approach adopted in this paper incorporates these structures to quantify systemic volatility by capturing higher-order dependencies across asset returns, thereby enriching the classical scalar notions of market risk.

\section{Financial Chaos Index}
\label{Sec:FCIX}

The FCIX is a tensor-based measure designed to quantify realized volatility in equity markets by capturing the mutual fluctuations among asset prices~\citep{ataei2021theory}. Unlike traditional scalar measures of volatility, the FCIX is constructed through higher-order tensorial representations, allowing for a richer depiction of systemic interdependencies.

To define the FCIX, consider a collection of $N$ assets and let $\mathbf{r}^{(t)} \in \mathbb{R}_{>0}^N$ denote the vector of rates of return at time $t \in \mathcal{T}$, where each entry is computed as
\begin{equation}
	r^{(t)}_i = \frac{C^{(t)}_i}{C^{(t-1)}_i},
\end{equation}
where $C^{(t)}_i$ represents the adjusted closing price of asset $i$ at time $t$. In our empirical analysis, we consider $N = 811$ assets, selected as those that have appeared at any time in the S\&P 500 index from January 1990 to December 2023 and for which no missing price data were observed.\footnote{Price data were obtained from the Center for Research in Security Prices (CRSP) via the Wharton Research Data Services (WRDS).} The time index set $\mathcal{T}$ comprises $|\mathcal{T}| = 8265$ daily and $|\mathcal{T}| = 408$ monthly observations, respectively.

At each time step $t$, we construct a {reciprocal pairwise comparison matrix} (RPCM) $\mathbf{A}^{(t)} \in \mathbb{R}_{>0}^{N \times N}$ via the relation
\begin{equation}
	\mathbf{A}^{(t)} = \mathbf{r} \circ \left(\mathbf{r}^{-1}\right)^\intercal,
\end{equation}
where the notation $\mathbf{r}^{-1}$ denotes the element-wise reciprocal of the return vector $\mathbf{r}^{(t)}$. Each entry $A_{ij}^{(t)}$ thus represents the relative strength of asset $i$ compared to asset $j$ at time $t$, a structure often referred to as an {advantage matrix}; see, e.g.,~\cite{ataei2020time}.

\begin{figure}
	\centering
	\includegraphics[scale=0.2]{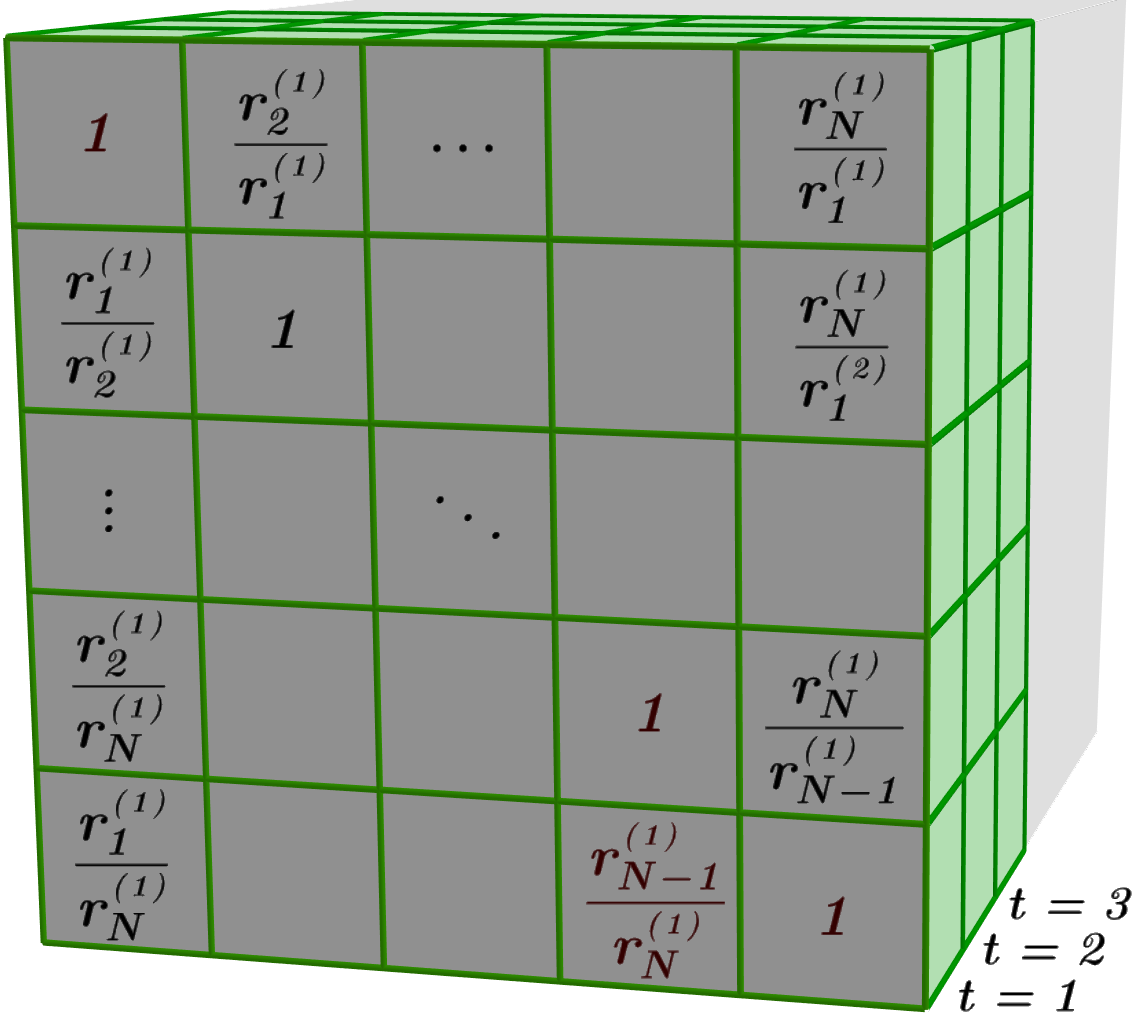}
	\caption{Schematic illustration of a reciprocal pairwise comparison tensor (RPCT)~\citep{ataei2021theory}.}
	\label{Fig:RPCT}
\end{figure}

By stacking these matrices across time, we construct a third-order tensor $\boldsymbol{\mathcal{A}} \in \mathbb{R}_{>0}^{N \times N \times |\mathcal{T}|}$, referred to as the {reciprocal pairwise comparison tensor} (RPCT), as illustrated schematically in Figure~\ref{Fig:RPCT}.

The core of the FCIX construction lies in approximating the RPCT by a rank-one tensor. Specifically, we solve the constrained optimization problem
\begin{equation}
	\begin{aligned}
		\min_{\tilde{\mathbf{x}}, \tilde{\mathbf{y}}, \tilde{\mathbf{z}}} \quad & \left\| \boldsymbol{\mathcal{A}} - \tilde{\mathbf{z}} \circ \left( \tilde{\mathbf{x}} \circ \tilde{\mathbf{y}}^\intercal \right) \right\|_F \\
		\text{s.t.} \quad & \tilde{z}_t \left( \tilde{\mathbf{x}} \circ \tilde{\mathbf{y}}^\intercal \right) > \mathbf{0}_N, \quad t = 1, \ldots, |\mathcal{T}|, \\
		& \tilde{\mathbf{x}}, \tilde{\mathbf{y}} \in \mathbb{R}_{>0}^N, \quad \tilde{\mathbf{z}} \in \mathbb{R}_{\geq 0}^{|\mathcal{T}|},
	\end{aligned}
\end{equation}
where $\mathbf{0}_N$ is the $N \times N$ zero matrix. The constraint ensures that all elements remain positive, preserving the multiplicative structure inherent in return ratios.

Once the rank-one approximation $\tilde{\boldsymbol{\mathcal{A}}} = \tilde{\mathbf{z}} \circ \left( \tilde{\mathbf{x}} \circ \tilde{\mathbf{y}}^\intercal \right)$ is obtained, the FCIX at time $t$ is defined via an inconsistency measure based on the dominant eigenvalue of the $t$th frontal slice of $\tilde{\boldsymbol{\mathcal{A}}}$. Specifically,
\begin{equation}
	\operatorname{FCIX}(t) := \psi(t) = \frac{\lambda_{\max}^{(t)} - N}{N-1},
\end{equation}
where $\lambda_{\max}^{(t)}$ denotes the largest eigenvalue of the matrix corresponding to the $t$th slice.

As demonstrated in~\cite{ataei2021theory}, there exists an elegant correspondence between the polyadic decomposition of RPCTs and the spectral properties of their constituent RPCMs, implying that the scaling vector $\tilde{\mathbf{z}}$ can be interpreted as a surrogate for the FCIX series, up to a positive rescaling.

Thus, the FCIX encapsulates the degree of mutual inconsistency and dispersion among asset returns in a manner that is sensitive to both local fluctuations and higher-order systemic co-movements. It offers a mathematically rigorous, geometrically interpretable, and computationally tractable measure of market volatility.

\begin{figure}
	\centering
	\includegraphics[width=\textwidth]{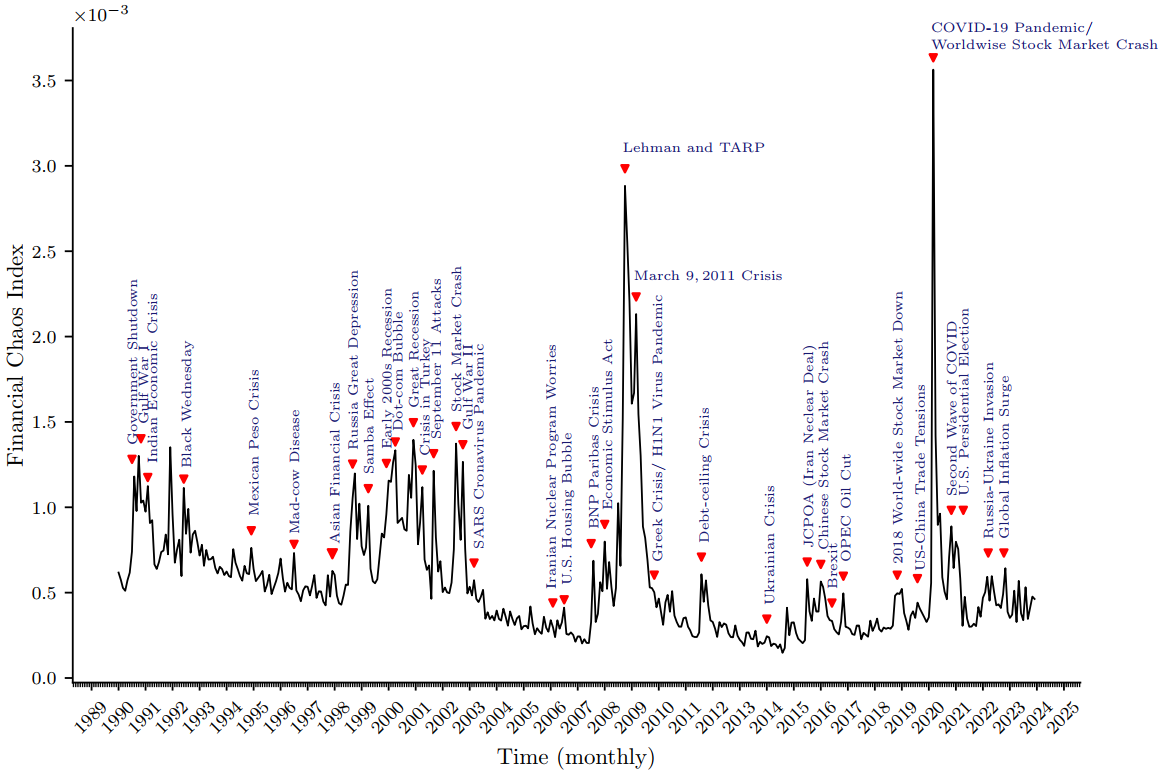} 
	\caption[Monthly $\mathrm{FCIX}$ during January $1990$-December $2023$.]{Monthly $\mathrm{FCIX}$ during January $1990$-December $2023$.} 
	\label{Fig:MFCI_Annotated} 
\end{figure}

A longitudinal inspection of monthly FCIX realizations from January 1990 to December 2023 reveals substantial temporal heterogeneity, undermining the hypothesis of stationary or smoothly evolving market dynamics. As shown in Figure~\ref{Fig:MFCI_Annotated}, the FCIX exhibits intermittent surges and contractions, often coinciding with major financial, economic, and geopolitical disruptions. Episodes such as the Dot-com bubble (2000--2002), the Global Financial Crisis (2008--2009), and the COVID-19 pandemic (2020) are each marked by pronounced elevations in the index, indicating acute systemic stress and structural dislocation.

The empirical behavior of the FCIX displays a notable asymmetry. Transitions to high-volatility regimes are typically abrupt and crisis-driven, while reversions to lower-volatility states occur more gradually. This asymmetry reflects well-documented phenomena in financial markets, wherein negative information triggers faster market responses than stabilization or recovery. The clustering of elevated FCIX values following major crises is consistent with volatility persistence and reinforces the view that financial stress tends to propagate over extended periods.

The dynamic behavior of the FCIX motivates a regime-switching interpretation of financial market evolution. In traditional asset pricing models, asset returns are often modeled as realizations from stochastic processes with constant mean and variance~\citep{campbell1999force, campbell2000asset}. However, the FCIX suggests an alternative perspective, wherein realized volatility transitions among discrete regimes characterized by distinct statistical properties. These transitions are frequently precipitated by fundamental shocks to the macro-financial environment, such as monetary policy realignments, sovereign debt crises, technological innovations, or geopolitical upheavals. Such shocks propagate through liquidity channels, reshape investor expectations, and alter perceived risk premia, effects that are subsequently encoded in the time series of the FCIX.

To formally investigate these regime transitions, we apply retrospective change-point detection methods to the FCIX time series. The literature on time series segmentation by change-point techniques is extensive; e.g., see \cite{truong2020selective} for the most recent review. Our methodological approach is nonparametric and model-free, operating on a high-dimensional mapping of the empirical distribution of the FCIX. Specifically, we implicitly embed the time series into a reproducing kernel Hilbert space defined by a suitable kernel function, thereby enabling change-point detection without presupposing a particular form for the underlying data-generating process. This kernel-based segmentation approach offers several advantages: it flexibly captures nonlinear dependencies, accommodates complex data structures, and remains robust to heavy-tailed distributions. Detailed theoretical underpinnings for such procedures can be found in \cite{ataei2021theory,arlot2019kernel, desobry2005online, harchaoui2007retrospective, harchaoui2008kernel}. By incorporating this model-free framework, we segment the FCIX into intervals of internal statistical homogeneity, thereby identifying latent regime shifts without imposing restrictive distributional assumptions.

\begin{figure}
	\centering
	\includegraphics[width=\textwidth]{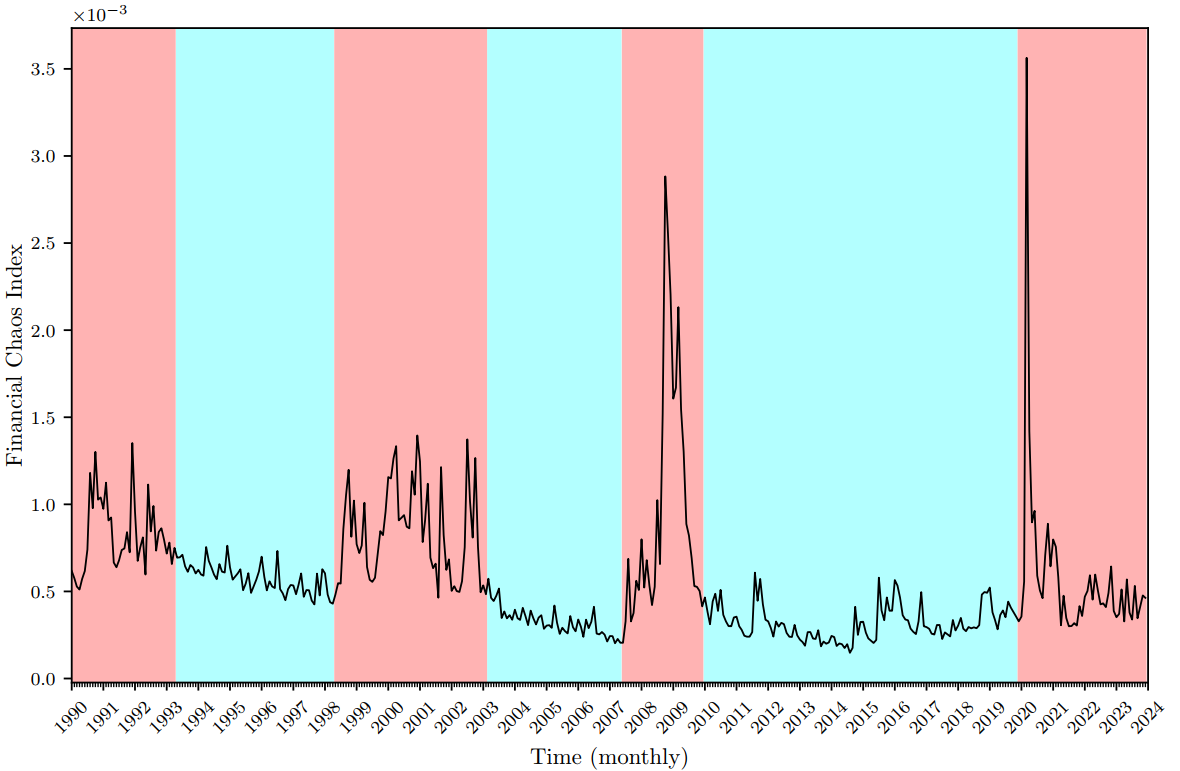} 
	\caption[Stock market segmentation during January $1990$-December $2023$.]{Stock market segmentation during January $1990$-December $2023$.} 
	\label{Fig:Segmentation} 
\end{figure}

The resulting segmentation is illustrated in Figure~\ref{Fig:Segmentation}. Several prominent breakpoints emerge from the analysis, including regime transitions associated with the 1997 Asian Financial Crisis, the onset of the 2008 Global Financial Crisis, the European Sovereign Debt Crisis during the early 2010s, and the COVID-19 pandemic in 2020. These regime shifts are not isolated anomalies but rather reflect profound structural reconfigurations in market dynamics. The persistent elevation of FCIX values following these inflection points indicates prolonged phases of systemic vulnerability and amplified risk perception.

The foregoing analysis substantiates the hypothesis that financial markets are intrinsically nonstationary systems, whose dynamics evolve in response to continuous informational shocks, policy interventions, and macroeconomic shifts. The segmentation of the FCIX into statistically distinct regimes, along with the heavy-tailed distributions observed within these regimes, supports a regime-switching framework for understanding market behavior. This framework has critical implications for risk management and financial stability monitoring. Models that ignore the presence of discrete regime shifts risk underestimating systemic vulnerability and tail risks. By contrast, models that incorporate dynamic regime identification, such as those incorporating the FCIX, enable more robust forecasting and offer early warning signals of evolving financial instability. Such capabilities equip policymakers and financial institutions with the tools necessary for preemptive intervention, resilience building, and the mitigation of systemic risk in an increasingly complex global financial system.

\begin{figure}
	\centering
	\includegraphics[width=\textwidth]{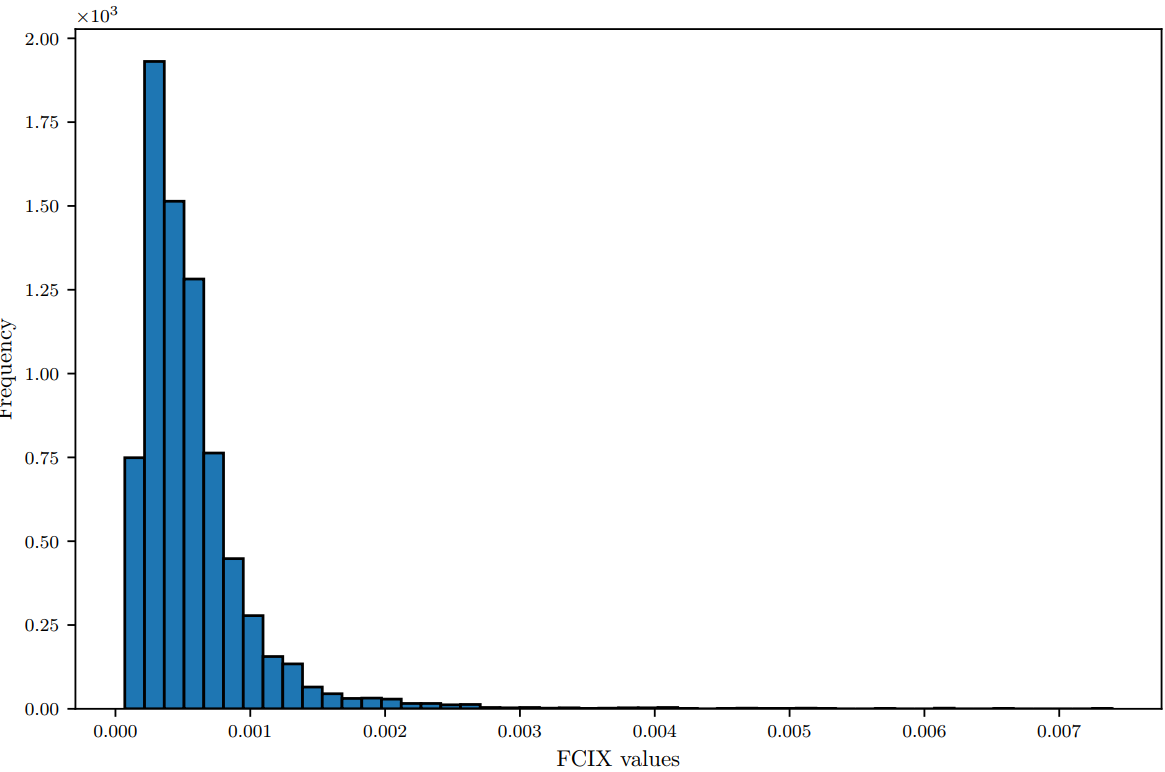}
	\caption[Frequency histogram for daily $\mathrm{FCI}_t$.]{Frequency histogram for $\mathrm{FCIX}$ during January $1990$-December $2023$.} 
	\label{Fig:Histogram}
\end{figure}

\section{Regime Analysis}
\label{Sec:Regime}
In this section, we conduct a statistical examination of the regime structure underlying the stock market's dynamic behavior. As a preliminary step, we analyze the empirical distribution of FCIX realizations. As illustrated in Figure~\ref{Fig:Histogram}, the histogram exhibits a markedly heavy-tailed shape, a hallmark of complex systems operating far from equilibrium. The majority of FCIX values are concentrated in the low-to-moderate range, corresponding to periods of relative market stability. However, the presence of an elongated right tail reflects the occasional but impactful episodes of systemic disruption, such as financial crises, during which the FCIX surges dramatically.

This dual behavior suggests that the FCIX is governed by a hybrid distributional structure. The clustering around moderate values is consistent with a lognormal model, which captures the compounding effects of numerous small, independent fluctuations. In contrast, the empirical rarity yet disproportionate magnitude of extreme FCIX spikes aligns more closely with power-law behavior, an observation frequently encountered in self-organizing critical systems and financial time series exhibiting tail dependence. This coexistence of lognormal bulk and power-law tail supports the hypothesis that financial markets alternate between phases of constrained stochastic fluctuation and episodic bursts of disorder, underscoring the relevance of regime-based modeling.

To accommodate both the routine dynamics and extreme deviations observed in FCIX realizations, we employ the MLP distribution \citep{basu2004power,basu2015mlp}. This model integrates two distinct behaviors within a unified statistical framework: lognormality governs the distribution's core, corresponding to low and moderate FCIX values, while a power-law tail emerges at higher values, capturing the heavy-tailed behavior associated with systemic crises. The lognormal regime reflects the cumulative effect of small, multiplicative shocks prevalent during periods of market stability, whereas the power-law tail encodes the disproportionate impact of rare, large-scale disruptions such as financial collapses or geopolitical dislocations.

The transition from lognormal to power-law behavior reflects the latent fragility of financial markets, systems which, while appearing stable under normal conditions, remain susceptible to endogenous amplification or exogenous shocks capable of precipitating critical transitions. In contrast to conventional parametric models that systematically underrepresent tail risk, the MLP offers a more expressive and empirically consistent representation of financial volatility. Its functional adaptability facilitates more accurate stress testing and scenario analysis, particularly in forecasting the likelihood and severity of extreme events. Moreover, the MLP's sensitivity to deviations from its expected structure enables its deployment as a diagnostic tool for early warning, offering a quantitative basis for preemptive policy or portfolio adjustments aimed at enhancing systemic resilience.

Let $\Psi_t$ denote a stochastic process associated with the value of $\mathrm{FCIX}$ at time $t$, with its finite realization vector represented as $\boldsymbol{\psi} = [\psi_1, \psi_2, \dots, \psi_{\mathcal{T}}]^\intercal$. The MLP distribution arises naturally from the temporal evolution of $\Psi_t$ under the hypothesis that the growth duration of its realizations is itself a random variable. Specifically, suppose that an initial quantity $\psi_{t_0}$ is drawn from a lognormal distribution at time $t = t_0$, parameterized by location $\mu$ and shape $\sigma$. The subsequent growth of $\Psi_t$ for $t > t_0$ is then governed by a deterministic exponential mechanism as follows:
\begin{equation}
	\frac{\partial \Psi_t}{\partial t} = \kappa \Psi_t,
\end{equation}
where $\kappa > 0$ denotes the \textit{growth rate}. This leads to a temporal shift in the lognormal mean, such that after a fixed duration $\Delta t$, the distribution remains lognormal with updated location parameter $\mu + \kappa \Delta t$, while retaining the original dispersion $\sigma$.

To encode uncertainty in the duration of this growth phase, we further model the elapsed time $\Delta t$ as a continuous random variable governed by an exponential distribution with \emph{stopping rate} parameter $\zeta > 0$. That is, the probability density function for the random growth duration is given by
\begin{equation}
	f(t) = \zeta \exp(-\zeta t), \qquad t \geq 0.
\end{equation}
Under this formulation, the MLP distribution emerges as a statistical framework wherein a lognormal base distribution is convolved with an exponential stopping-time mechanism, thereby producing a distribution whose body is governed by multiplicative stochasticity and whose tail exhibits power-law scaling. This provides a principled stochastic foundation for the hybrid structure observed in empirical FCIX realizations.

The probability density function associated with the random variable $\Psi_t$ under the MLP model is given by
\begin{equation}
	\label{Eq:Distr_Log_N}
	g(\psi_t;\mu,\sigma,\omega) = \left(\dfrac{\omega}{2}\right) \left(\dfrac{1}{\psi_t}\right)^{1+\omega} \exp\left\{ \omega \mu + \dfrac{\omega^2 \sigma^2}{2} \right\} \mathrm{erfc}\left\{ \frac{1}{\sqrt{2}} \left( \omega\sigma - \frac{\log \psi_t - \mu}{\sigma} \right) \right\},
\end{equation}
where $\psi_t > 0$, $\mu \in \mathbb{R}$, $\sigma > 0$, and $\omega = \zeta / \kappa > 0$ is the \textit{power-law tail index} controlling the heaviness of the distribution's upper tail. Here, $\mathrm{erfc}(\cdot)$ denotes the complementary error function. The corresponding cumulative distribution function is given by
\begin{equation}
	\resizebox{1\hsize}{!}  { $G(\psi_t;\mu,\sigma,\omega) = \dfrac{1}{2} \mathrm{erfc}\left\{ - \dfrac{\log \psi_t - \mu}{\sqrt{2}\sigma} \right\}  - \dfrac{1}{2\psi_t^\omega}  \exp\left\{ \omega \mu + \dfrac{\omega^2 \sigma^2}{2} \right\} \mathrm{erfc}\left\{ \dfrac{1}{\sqrt{2}} \left( \omega\sigma - \dfrac{\log \psi_t - \mu}{\sigma} \right) \right\}$ }.
\end{equation}

This formulation reveals the hybrid structure of the MLP distribution. In the regime $\psi_t \to 0$, the first term in $G(\psi_t;\mu,\sigma,\omega)$ dominates, and the distribution converges to a lognormal form with parameters $(\mu, \sigma)$. This captures the behavior of small-to-moderate FCIX values arising from routine market fluctuations. Conversely, in the limit $\sigma \to 0$, the MLP asymptotically approaches a pure power-law form, governed entirely by the tail index $\omega$. This characterizes the heavy-tailed regime associated with extreme market disruptions. Hence, the MLP family encapsulates both statistical regularity and disorder within a single parametrized structure, offering analytical flexibility in modeling the empirical distribution of FCIX realizations across regimes.

The MLP distribution admits closed-form expressions for its first two moments under suitable conditions on the tail index parameter $\omega$. The expected value of $\Psi_t$ is given by
\begin{equation}
	\mathbb{E}[\Psi_t] = \omega \exp\left\{ 2\mu + \sigma^2 \right\} \left( \frac{\exp\left\{\sigma^2 \right\} }{\omega - 2} - \frac{\omega}{\left( \omega - 1 \right)^2} \right), \qquad \omega > 2,
\end{equation}
while the variance is expressed as
\begin{equation}
	\mathrm{Var}[\Psi_t] = \frac{\omega}{\omega - 1} \exp\left\{ \mu + \frac{\sigma^2}{2} \right\}, \qquad \omega > 1.
\end{equation}

These moment conditions imply that the existence of finite variance requires $\omega > 1$, while a finite mean requires the stricter constraint $\omega > 2$. This demarcation highlights the heavy-tailed nature of the MLP distribution: as $\omega$ approaches its lower bounds, the moments diverge, reflecting the increasing contribution of rare but extreme values to the overall distribution. Such behavior underscores the necessity of careful tail-index estimation in practical applications, particularly in the context of risk quantification and scenario-based forecasting where moment stability is critical.

To model the presence of multiple latent market regimes, we posit that the FCIX realizations arise from a finite mixture of $R$ distinct  MLP components. The number of regimes $R \geq 1$ is unknown a priori and is to be inferred from the data. Let $\boldsymbol{\theta}_r = [\mu_r, \sigma_r, \omega_r, \pi_r]$ denote the parameter vector associated with regime $r$, where $\pi_r$ represents the regime's mixing proportion (i.e., the relative frequency of observations attributed to regime $r$). The global distribution of $\Psi_t$ is then modeled by the following mixture density
\begin{subequations}
	\label{Eq:Mixture_Density}
	\begin{align}
		h(\psi_t;\boldsymbol{\theta}) &= \sum_{r=1}^{R} \pi_r \, g(\psi_t;\boldsymbol{\theta}_r)\\
		&= \sum_{r=1}^{R} \pi_r \left(\frac{\omega_r}{2}\right) \left(\frac{1}{\psi_t}\right)^{1+\omega_r} \exp\left\{ \omega_r \mu_r + \frac{\omega_r^2 \sigma_r^2}{2} \right\} 
		\mathrm{erfc}\left\{ \frac{1}{\sqrt{2}} \left( \omega_r\sigma_r - \frac{\log \psi_t - \mu_r}{\sigma_r} \right) \right\},
	\end{align}
\end{subequations}
where $g(\psi_t;\boldsymbol{\theta}_r)$ denotes the MLP density for the $r$th regime. The full parameter space is captured by the matrix
\begin{equation}
	\label{Eq:Theta_Mat}
	\boldsymbol{\theta} = 
	\begin{pmatrix}
		\mu_1 & \sigma_1 & \omega_1 & \pi_1 \\
		\mu_2 & \sigma_2 & \omega_2 & \pi_2 \\
		\vdots & \vdots & \vdots & \vdots \\
		\mu_R & \sigma_R & \omega_R & \pi_R \\
	\end{pmatrix},
\end{equation}
with $\sum_{r=1}^{R} \pi_r = 1$ and $\pi_r > 0$ for all $r$.

The mixture model in \eqref{Eq:Mixture_Density} represents a flexible model of the empirical FCIX distribution, with each regime corresponding to a distinct hypothesis in the model's alternative component specification \citep{hamilton1994time}. Estimation of $\boldsymbol{\theta}$ and inference on the true value of $R$ can be performed via likelihood-based methods. This mixture framework enables both regime identification and the characterization of heterogeneity in market dynamics, offering a data-driven foundation for capturing hidden structural shifts in financial volatility.

To operationalize estimation of the mixture model parameters, we discretize the observed realizations of $\Psi_t$ into $L$ non-overlapping intervals, forming a binned representation of the empirical distribution. Let $|\mathcal{T}|$ denote the total number of observed FCIX values, which are categorized into $L$ bins $\{\mathit{B}_l\}_{l=1}^{L}$. The bin boundaries are defined by a sequence $\{b_l\}_{l=1}^{L}$ such that
\begin{equation}
	\mathit{B}_l = \left[b_l - \frac{b_l - b_{l-1}}{2},\ b_l + \frac{b_{l+1} - b_l}{2} \right], \quad l = 2,\dots,L-1,
\end{equation}
with the boundary cases specified by
\begin{align}
	\mathit{B}_1 &= (0,\, b_1 + \tfrac{b_2 - b_1}{2}], \\
	\mathit{B}_L &= [b_L - \tfrac{b_L - b_{L-1}}{2},\, \infty),
\end{align}
where $b_1 = \min\{\psi_t\}$ and $b_L = \max\{\psi_t\}$. The number of bins $L$ is determined via the Rice rule, yielding
\begin{equation*}
	L = 2 |\mathcal{T}|^{1/3} = 41,
\end{equation*}
which provides a heuristic balance between resolution and robustness in histogram-based estimation.

Let $\hat{p}_l$ denote the observed proportion of data falling into bin $\mathit{B}_l$, and let $p_l(\boldsymbol{\theta})$ denote the corresponding theoretical probability under the mixture model given by
\begin{align}
	p_l(\boldsymbol{\theta}) &= \mathbb{P}[\Psi_t \in \mathit{B}_l \mid \boldsymbol{\theta}] \\
	&= \int_{\inf(\mathit{B}_l)}^{\sup(\mathit{B}_l)} h(\psi_t; \boldsymbol{\theta})\, d\psi_t \nonumber \\
	&= \sum_{r=1}^{R} \pi_r \int_{\inf(\mathit{B}_l)}^{\sup(\mathit{B}_l)} g(\psi_t; \boldsymbol{\theta}_r)\, d\psi_t \\
	&= \sum_{r=1}^{R} \pi_r \left[ G(\sup(\mathit{B}_l); \boldsymbol{\theta}_r) - G(\inf(\mathit{B}_l); \boldsymbol{\theta}_r) \right],
\end{align}
where $G(\cdot;\boldsymbol{\theta}_r)$ is the cumulative distribution function of the MLP density associated with regime $r$.

Assuming the binned counts follow a multinomial distribution, the likelihood function for the vector of observed proportions $\hat{\mathbf{p}} = [\hat{p}_1, \hat{p}_2, \dots, \hat{p}_L]^\intercal$ is given by
\begin{equation}
	\label{Eq:Likelihood_Fun}
	\mathcal{L}(\boldsymbol{\theta} \mid \hat{\mathbf{p}}) = \prod_{l=1}^{L} \left( p_l(\boldsymbol{\theta}) \right)^{|\mathcal{T}|  \hat{p}_l}.
\end{equation}
Taking the logarithm of the multinomial likelihood in \eqref{Eq:Likelihood_Fun}, the log-likelihood function takes the form
\begin{equation}
	\label{Eq:Log_L}
	\log \mathcal{L}(\boldsymbol{\theta} \mid \hat{\mathbf{p}}) = |\mathcal{T}| \sum_{l=1}^{L} \hat{p}_l \log p_l(\boldsymbol{\theta}).
\end{equation}

To facilitate comparison between the empirical proportions $\hat{p}_l$ and their model-based counterparts $p_l(\boldsymbol{\theta})$, we subtract the empirical entropy term $|\mathcal{T}| \sum_{l=1}^{L} \hat{p}_l \log \hat{p}_l$, which is constant with respect to the model parameters, and multiply the resulting expression by a factor of $2$, yielding
\begin{equation}
	\log \mathcal{L}(\boldsymbol{\theta} \mid \hat{\mathbf{p}}) = 
	\begin{cases}
		2|\mathcal{T}| \sum\limits_{l=1}^{L} \hat{p}_l \log \left[ \dfrac{p_l(\boldsymbol{\theta})}{\hat{p}_l} \right], & \text{if } \hat{p}_l > 0, \\
		0, & \text{if } \hat{p}_l = 0.
	\end{cases}
\end{equation}
This expression is equivalent to a scaled version of the Kullback-Leibler divergence between the empirical distribution $\hat{\mathbf{p}}$ and the model-implied distribution $p(\boldsymbol{\theta})$. Specifically,
\begin{equation}
	\label{Eq:KL_Div}
	\log \mathcal{L}(\boldsymbol{\theta} \mid \hat{\mathbf{p}}) = -2|\mathcal{T}| \sum_{l=1}^{L} D_{\mathrm{KL}} \left( \hat{p}_l \,\|\, p_l(\boldsymbol{\theta}) \right),
\end{equation}
where $D_{\mathrm{KL}}( \hat{p}_l \,\|\, p_l ) = \hat{p}_l \log \left( \hat{p}_l / p_l \right)$ for $\hat{p}_l > 0$.

When parameter estimates are treated as random variables, the right-hand side of \eqref{Eq:KL_Div} approximates a $\chi^2$ distribution with $L - 4R - 1$ degrees of freedom, reflecting the effective dimensionality reduction due to estimation of $4R$ parameters in the MLP mixture model (three for distributional shape and one for the mixing proportion per regime). This statistical interpretation enables hypothesis testing, model validation, and asymptotic comparison among competing mixture configurations.

To estimate the parameter matrix $\boldsymbol{\theta}$ for a fixed number of regimes $R$, we solve the following constrained optimization problem
\begin{subequations}
	\label{Eq:Regime}
	\begin{alignat}{3}
		\max_{\substack{\boldsymbol{\mu},\boldsymbol{\sigma},\boldsymbol{\omega},\boldsymbol{\pi}}} \,\,&   \log \mathcal{L}(\boldsymbol{\theta} \mid \hat{\mathbf{p}}) & \\
		\mathrm{s.t.} \quad & \sum_{r=1}^{R} \pi_r = 1, & \\
		& \mu_r \in (-\infty, \infty), \quad & r=1,\dots,R, \\
		& \sigma_r \in (0, \infty), \quad & r=1,\dots,R, \\
		& \omega_r \in (0, \infty), \quad & r=1,\dots,R, \\
		& \pi_r \in (0, 1], \quad & r=1,\dots,R,
	\end{alignat}
\end{subequations}
where $\boldsymbol{\mu} = [\mu_1, \dots, \mu_R]^\intercal$, $\boldsymbol{\sigma} = [\sigma_1, \dots, \sigma_R]^\intercal$, $\boldsymbol{\omega} = [\omega_1, \dots, \omega_R]^\intercal$, and $\boldsymbol{\pi} = [\pi_1, \dots, \pi_R]^\intercal$ represent the vectors of location, scale, tail index, and mixing proportions, respectively.

Model adequacy for each hypothesized value of $R$ is evaluated via the divergence-based log-likelihood function defined in \eqref{Eq:KL_Div}, which, under regularity conditions, asymptotically follows a chi-squared distribution. The null hypothesis that the model with $R$ regimes provides a satisfactory fit is retained if the corresponding $p$-value exceeds a chosen significance level (e.g., $\alpha = 0.05$), thereby avoiding overfitting through superfluous components.

In practice, the true number of regimes is unknown and must be inferred empirically. To that end, we conduct goodness-of-fit tests sequentially for each candidate value of $R$ by solving the optimization problem \eqref{Eq:Regime} and computing the corresponding test statistic. Table~\ref{Regime_Results} reports the results of this regime analysis for the FCIX time series from January 1990 to December 2023. The underlying models were estimated using a trust-region Newton-conjugate-gradient algorithm, which offers both numerical stability and convergence efficiency for high-dimensional, non-convex likelihood landscapes.

As shown in Table~\ref{Regime_Results}, both the two-regime and three-regime mixture models are statistically supported at all conventional levels of significance ($\alpha = 0.01,\ 0.05,\ 0.10$), while alternative models with either fewer or more components, including the one-regime specification corresponding to a random walk, fail the goodness-of-fit test. Among the admissible models, the three-regime formulation yields the lowest $\chi^2$ value and highest $p$-value, thereby offering the best compromise between model parsimony and explanatory adequacy.

Accordingly, we select the three-regime MLP mixture as the most appropriate regime-delegating paradigm for the distribution of FCIX values. This choice not only provides a statistically superior fit but also reinforces the empirical finding that FCIX realizations are heterogeneously distributed across distinct volatility regimes. The rejection of the single-regime hypothesis implies that the temporal structure of market volatility is not homogeneous, underscoring the existence of clustered phases of distinct systemic dynamics. This segmentation further corroborates earlier observations regarding structural shifts and regime persistence in the FCIX time series.
 
The parameter estimates for the three-regime MLP mixture indicate that the dominant regime, accounting for $54.8\%$ of the total mass, corresponds to an {intermediate-chaos} (IC) state, characterized by a mean FCIX level of $0.0006$ and a standard deviation of $0.0003$. The second most prevalent component, representing $35.5\%$ of the distribution, is identified as the {low-chaos} (LC) regime, with a mean of $0.0003$ and a standard deviation of $0.0001$. Notably, both LC and IC regimes exhibit relatively low dispersion, indicating that FCIX values within these states evolve in a more predictable and less volatile manner.

In contrast, the {high-chaos} (HC) regime, although contributing only $9.7\%$ to the mixture, is marked by a significantly elevated mean of $0.0012$ and a standard deviation of $0.0009$. This regime encapsulates periods of extreme systemic instability, as evidenced by its markedly higher variability. The disproportion between its statistical weight and volatility highlights its role as a low-frequency but high-impact state in market evolution.

\begin{figure}
	\centering
	\resizebox{1\textwidth}{0.4\textheight}{\input{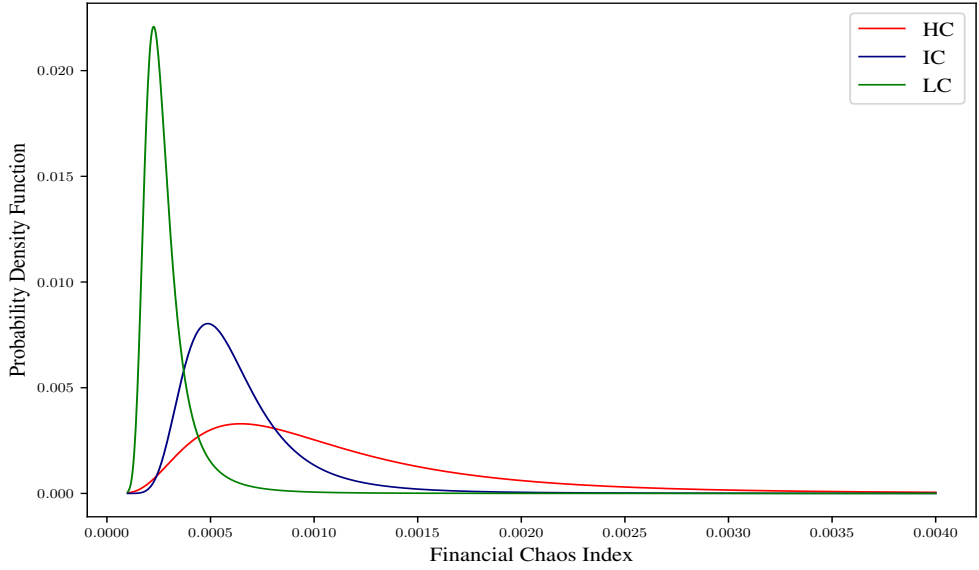}}
	\caption{Plots of the probability density functions of market regimes.}
	\label{fig_Regimes}
\end{figure}

Figure~\ref{fig_Regimes} illustrates the estimated probability density functions associated with each inferred regime, further underscoring the multimodal structure of the FCIX distribution and the contrast in scale and variability across regimes.

\begin{table}
	\centering 
	\small
	\caption[Results of regime analysis for daily FCIX.]{Results of regime analysis for daily FCIX during January 1990-December 2023.}
	\label{Regime_Results}
	\begin{tabular}{crrrrr} 
		\toprule
		& 1-regime & 2-regime & 3-regime & 4-regime & 5-regime  \\
		\midrule
		$\mu_1$ & -7.988 & -8.414 & -8.502 & -8.287 & -8.405 \\
		$\mu_2$ & --     & -7.646 & -7.697 & -7.949 & -7.796 \\
		$\mu_3$ & --     & --     & -7.204 & -7.883 & -7.446 \\
		$\mu_4$ & --     & --     & --     & 6.397  & -7.057 \\
		$\mu_5$ & --     & --     & --     & --     & -6.296 \\ \\
		$\sigma_1$ & 0.525 & 0.306 & 0.218 & 0.479 & 0.091 \\
		$\sigma_2$ & --    & 0.245 & 0.308 & 0.535 & 0.204 \\
		$\sigma_3$ & --    & --    & 0.573 & 0.519 & 0.213 \\
		$\sigma_4$ & --    & --    & --    & 0.922 & 0.295 \\
		$\sigma_5$ & --    & --    & --    & --    & 0.421 \\ \\
		$\omega_1$ & 3.159 & 2.662 & 3.584 & 2.131 & 0.091 \\
		$\omega_2$ & --    & 2.627 & 3.837 & 4.593 & 0.204 \\
		$\omega_3$ & --    & --    & 3.869 & 3.644 & 0.213 \\
		$\omega_4$ & --    & --    & --    & 2.292 & 0.295 \\
		$\omega_5$ & --    & --    & --    & --    & 0.421 \\ \\
		$\pi_1$    & 1.000 & 0.522 & 0.355 & 0.246 & 0.039 \\
		$\pi_2$    & --    & 0.478 & 0.548 & 0.093 & 0.206 \\
		$\pi_3$    & --    & --    & 0.097 & 0.651 & 0.225 \\
		$\pi_4$    & --    & --    & --    & 0.010 & 0.245 \\
		$\pi_5$    & --    & --    & --    & --    & 0.285 \\ \\
		$\chi^2$   & 96.743 & 24.249 & 18.594 & 61.310 & 73.663 \\
		$p$-value  & 0.000  & 0.983  & 0.995  & 0.002  & 0.000 \\
		\bottomrule
	\end{tabular}
\end{table}

To infer the prevailing market regime at each point in time, we define a latent random variable \( S_t \) representing the state of the financial chaos index FCIX at time \( t \). The support of \( S_t \) comprises three distinct regimes: \( s_t \in \{ \mathrm{LC}, \mathrm{IC}, \mathrm{HC} \} \), corresponding to the low-chaos, intermediate-chaos, and high-chaos states, respectively.

The probability that a given observation \( \psi_t \) arises from regime \( s_t \), conditional on the observed value and the estimated parameters, is given by
\begin{equation}
	h(S_t = s_t \mid \psi_t; \boldsymbol{\theta}) = \frac{h_{s_t}(\psi_t; \boldsymbol{\theta}_{s_t})}{h(\psi_t; \boldsymbol{\theta})},
\end{equation}
where \( h_{s_t}(\psi_t; \boldsymbol{\theta}_{s_t}) \) denotes the regime-specific MLP density, and \( h(\psi_t; \boldsymbol{\theta}) \) is the global mixture density as defined in \eqref{Eq:Mixture_Density}.

This posterior formulation enables the classification of each FCIX realization \( \psi_t \) into its most likely generating regime, thereby producing a time-indexed sequence of dominant regime labels. Such a classification provides a granular view of market phase transitions and facilitates the detection of regime persistence, structural shifts, and volatility clustering across the observation horizon.

The temporal dynamics of market regimes can be further analyzed through a discrete-time Markov process, represented by a {transition probability matrix} \( \mathbf{P} \). Each entry \( P_{ij} \) denotes the estimated probability of transitioning from regime \( i \) to regime \( j \) over a single time increment. Letting the rows correspond to the current regime and the columns to the subsequent regime, the estimated transition matrix is given by
\[
\mathbf{P} = 
\begin{pmatrix}
	0.483 & 0.493 & 0.024 \\
	0.409 & 0.565 & 0.027 \\
	0.500 & 0.470 & 0.030
\end{pmatrix},
\]
where the rows and columns are indexed by the regime states \( \{\mathrm{LC}, \mathrm{IC}, \mathrm{HC}\} \).

The transition matrix \( \mathbf{P} \) offers a compact representation of regime persistence and inter-regime mobility within the FCIX framework. Diagonal entries reflect the probability of remaining in the same regime across consecutive time steps. Specifically, the low-chaos regime exhibits moderate persistence with \( P_{\mathrm{LC}, \mathrm{LC}} = 0.483 \), while the intermediate-chaos regime displays a higher stability level with \( P_{\mathrm{IC}, \mathrm{IC}} = 0.565 \). This suggests that moderate-volatility phases are relatively enduring and often serve as transitional buffers between stable and crisis states. In stark contrast, the high-chaos regime is markedly transient, with \( P_{\mathrm{HC}, \mathrm{HC}} = 0.030 \), indicating that periods of extreme disorder are typically short-lived. This aligns with empirical observations that financial crises, though severe, tend to dissipate once institutional or policy-based corrective mechanisms are activated.

The off-diagonal elements capture transitions between different volatility states. The probability of moving from LC to IC is 0.493, nearly equal to the probability of remaining in LC, illustrating the frequent oscillation between stability and moderate volatility, likely driven by macroeconomic adjustments and shifts in investor sentiment. Similarly, the probability of reverting from IC to LC is 0.409, reinforcing the tendency of markets to normalize following moderate stress. Transitions from LC and IC to HC occur infrequently (0.024 and 0.027, respectively), underscoring the rarity of extreme systemic disruptions. However, once the system enters the HC regime, the probabilities of transitioning back to LC (0.500) or IC (0.470) vastly exceed the probability of persistence, reflecting the market's inherent resilience and mean-reverting tendencies following crises.

The persistence of the intermediate-chaos regime further emphasizes its role as a stabilizing intermediary. It operates as a critical zone in which financial systems absorb shocks, recalibrate expectations, and potentially redirect trajectories toward stability or systemic stress. In contrast, the HC regime's brief duration underscores its episodic nature, emerging abruptly, generating concentrated instability, and then subsiding as structural corrections take hold. This dynamic interplay between regime states provides a lens through which to interpret the evolution of financial disorder and recovery across temporal horizons.

While the transition matrix characterizes short-term regime dynamics, understanding the long-term behavior of the market requires examining the {stationary distribution}. The stationary distribution vector \( \boldsymbol{\pi} \) represents the asymptotic probabilities of the system residing in each regime, independent of the initial state. It satisfies the equilibrium condition
\begin{equation}
	\boldsymbol{\pi} \mathbf{P} = \boldsymbol{\pi}, \qquad \text{subject to} \quad \sum_{i=1}^{3} \pi_i = 1,
\end{equation}
which ensures that, over an extended horizon, the net flow into and out of each regime remains balanced.

Based on the empirical transition matrix \( \mathbf{P} \), the stationary distribution is computed as follows:
\[
\boldsymbol{\pi} = 
\begin{pmatrix}
	0.444 \\
	0.530 \\
	0.026
\end{pmatrix},
\]
indicating that, in the long run, the system spends approximately 44.4\% of the time in the  LC regime, 53.0\% in the IC regime, and only 2.6\% in the HC regime.

The dominance of the IC regime in the stationary distribution highlights the market's natural tendency to fluctuate within a moderate volatility zone, shaped by ongoing macroeconomic adjustments and behavioral dynamics. The LC regime also maintains substantial long-run probability, reflecting the enduring presence of structurally stable periods in the financial system. In contrast, the HC regime has a vanishingly small stationary probability, consistent with the view that systemic crises are rare, episodic events. These findings complement the transition-based analysis and reinforce the notion that financial markets are largely governed by alternating episodes of stability and moderate disorder, punctuated by infrequent bursts of acute turbulence.

\section{Forces Driving Market Volatility}
\label{Sec:Forces}
This section investigates the external forces that shape stock market volatility, drawing on sentiment indices derived from economic, policy, financial and geopolitical news. Volatility measures are generally categorized into realized and implied forms. Realized volatility reflects ex post fluctuations in asset prices, and is captured here by the FCIX, which we have previously employed to detect structural regime transitions in historical market behavior. By contrast, implied volatility is a forward-looking measure derived from options markets, embodying investor expectations about future uncertainty. Among such indicators, the Chicago Board Options Exchange's VIX index remains the most prominent and widely referenced gauge of anticipated equity market volatility.

Focusing on implied rather than realized volatility in the present analysis is motivated by several considerations. First, implied volatility captures forward-looking risk perceptions, making it particularly sensitive to evolving macro-financial conditions. Second, as a derivative of option pricing models, the VIX reflects anticipatory concerns over hedging costs, tail risks, and the broader confidence environment. Third, substantial empirical evidence supports the VIX's predictive value for future realized volatility, liquidity stresses, and systemic fragility. Consequently, it serves not only as an uncertainty barometer but also as a latent indicator of financial instability.

To probe the informational underpinnings of implied volatility fluctuations, we construct predictive models for the one-step-ahead monthly value of the VIX, using sentiment-based predictors derived from the EMV tracker \cite{baker2019policy}. The EMV index aggregates article counts from major U.S. newspapers, focusing on those containing terms from three semantic domains: (i) economic or financial concepts, (ii) equity market references, and (iii) uncertainty-related language such as ``volatility,'' ``risk,'' or ``uncertainty.'' The EMV index is scaled to match the historical mean of the VIX and demonstrates strong empirical correlation with both realized and implied volatility measures, reinforcing its validity as a proxy for news-driven uncertainty.

Beyond the aggregate index, over forty category-specific EMV trackers have been developed, offering thematic granularity by quantifying the prevalence of macroeconomic news, fiscal and monetary policy debates, financial regulation, national security concerns, and trade policy developments. Notably, macroeconomic uncertainty features in 72\% of EMV articles, commodity markets in 44\%, fiscal policy in 35\%, and monetary policy in 30\%~\cite{baker2019policy}. This taxonomy enables a disaggregated exploration of how distinct uncertainty channels influence market expectations.

To estimate the relative importance of these thematic categories in driving implied volatility, we employ an elastic net regression framework. This regularized modeling approach accommodates multicollinearity among predictors and performs automatic variable selection, combining lasso (\( \ell_1 \)) and ridge (\( \ell_2 \)) penalties to balance sparsity and robustness. Importantly, the primary objective extends beyond mere forecasting accuracy to include inferential insights: by examining selected covariates and their coefficients across regimes, we aim to characterize the dynamic structure of market-relevant uncertainty.

This investigation is embedded within the broader regime-switching framework established through the FCIX segmentation. By linking sentiment-based predictors to forward-looking volatility measures, the analysis traces how diverse forms of uncertainty differentially impact market dynamics and investor anticipations across structural phases. In doing so, it provides empirical insights into whether regime shifts are predominantly precipitated by macroeconomic instability, policy dislocations, or geopolitical shocks. These findings bear direct relevance to financial stability surveillance and the development of early-warning systems for systemic risk.

To operationalize this analysis, we estimate a series of elastic net regression models across temporally segmented phases of financial market history. Specifically, we leverage the FCIX-driven segmentation of the time series into structurally homogeneous intervals, as depicted in Figure~\ref{Fig:Segmentation}. Each segment corresponds to a distinct historical episode marked by relative statistical stationarity and macro-financial coherence. The segmentation is summarized in Table~\ref{tab:segments}.

\begin{table}[h]
	\centering
	\caption{Segmented time periods.}
	\label{tab:segments}
	\begin{tabular}{cc}
		\toprule
		Segment Index & Time Period \\
		\midrule
		1 & 1990-01 -- 1993-06 \\
		2 & 1993-07 -- 1998-06 \\
		3 & 1998-07 -- 2002-12 \\
		4 & 2003-01 -- 2007-09 \\
		5 & 2007-10 -- 2009-12 \\
		6 & 2010-01 -- 2019-12 \\
		7 & 2020-01 -- 2023-12 \\
		\bottomrule
	\end{tabular}
\end{table}

Let \( \mathcal{T}_j \subset \mathcal{T} \) denote the set of monthly time indices associated with segment \( j \in \{1, 2, \dots, 7\} \). For each segment \( j \), we model the one-step-ahead value of the VIX as a linear function of category-specific EMV indices, denoted \( \mathbf{x}_t = [x_{1t}, \dots, x_{Kt}]^\intercal \in \mathbb{R}^K \), where each \( x_{kt} \) measures the scaled frequency of uncertainty-related news in category \( k \) at time \( t \).

The segment-specific elastic net model is given by
\begin{equation}
	{\mathrm{VIX}}_{t+1}^{(j)} = \beta_0^{(j)} + \sum_{k=1}^{K} \beta_k^{(j)} x_{kt} + \varepsilon_t^{(j)}, \qquad t \in \mathcal{T}_j,
\end{equation}
where the coefficients are estimated via the regularized optimization
\begin{equation}
	\widehat{\boldsymbol{\beta}}^{(j)} = \arg\min_{\boldsymbol{\beta}} \left\{ \dfrac{1}{|\mathcal{T}_j|} \sum\limits_{t \in \mathcal{T}_j} \left( \mathrm{VIX}_{t+1} - \beta_0^{(j)} - \sum_{k=1}^{K} \beta_k x_{kt} \right)^2 + \lambda^{(j)} \left[ \alpha^{(j)} \|\boldsymbol{\beta}\|_1 + \frac{1}{2}(1 - \alpha^{(j)}) \|\boldsymbol{\beta}\|_2^2 \right] \right\}.
\end{equation}
%\begin{equation}
%	\resizebox{1\hsize}{!}  { $\widehat{\boldsymbol{\beta}}^{(j)} = \arg\min_{\boldsymbol{\beta}} \left\{ \dfrac{1}{|\mathcal{T}_j|} \sum_{t \in \mathcal{T}_j} \left( \mathrm{VIX}_{t+1} - \beta_0^{(j)} - \sum_{k=1}^{K} \beta_k x_{kt} \right)^2 + \lambda^{(j)} \left[ \alpha^{(j)} \|\boldsymbol{\beta}\|_1 + \frac{1}{2}(1 - \alpha^{(j)}) \|\boldsymbol{\beta}\|_2^2 \right] \right\}.$ }
%\end{equation}
Here, \( \lambda^{(j)} \geq 0 \) controls the overall penalty strength, and \( \alpha^{(j)} \in [0, 1] \) governs the balance between lasso and ridge penalization. Both hyperparameters are tuned via cross-validation within each segment to optimize out-of-sample prediction performance.

This modeling framework allows the structure of information relevance to flexibly evolve over time, capturing temporal shifts in the thematic composition of uncertainty. By doing so, it sheds light on the differential roles played by macroeconomic, policy, and geopolitical drivers in shaping investor expectations under varying market regimes.

Table~\ref{tab:elasticnet-summary} summarizes the model diagnostics for each segment, while Table~\ref{tab:elasticnet-covariates} lists the most influential EMV predictors per segment, ordered by the magnitude of their coefficients. These results are interpreted in connection with the patterns of FCIX shown in Figure~\ref{Fig:Segmentation}, with particular emphasis on how the number, strength, and nature of explanatory forces evolve across regime transitions. It is well-established that implied and realized volatilities are tightly coupled, particularly during periods of systemic stress~\cite{ataei2021theory}. Consequently, patterns observed in the predictive structure of implied volatility also offer valuable insights into the informational environment governing realized market behavior, as elucidated in the segmentation analysis that follows.

\begin{table}
	\centering
	\caption{Performance summary of elastic nets.}
	\label{tab:elasticnet-summary}
	\begin{tabular}{cccccc}
		\toprule
		Segment & Time Period & \( \alpha \) & \( R^2 \) & MSE & RMSE \\
		\midrule
		1 & 1990--1993 & 0.10 & 0.709 & 0.0216 & 0.1469 \\
		2 & 1993--1998 & 0.60 & 0.638 & 0.0222 & 0.1490 \\
		3 & 1998--2002 & 0.75 & 0.691 & 0.0296 & 0.1721 \\
		4 & 2003--2007 & 0.85 & 0.696 & 0.0194 & 0.1393 \\
		5 & 2007--2009 & 0.80 & 0.917 & 0.0154 & 0.1242 \\
		6 & 2010--2019 & 0.45 & 0.702 & 0.0185 & 0.1359 \\
		7 & 2020--2023 & 0.70 & 0.854 & 0.0243 & 0.1559 \\
		\bottomrule
	\end{tabular}
\end{table}

\begin{table}
	\centering
	\caption{Dominant drivers of implied volatility (VIX) across regimes.}
	\label{tab:elasticnet-covariates}
	\begin{tabular}{cccccc}
		\toprule
		Segment  & Predictor 1 & Predictor 2 & Predictor 3 & Predictor 4 & Predictor 5 \\
		\midrule
		1 & Disease (-0.63) & Labor (-0.61) & Commodities (+0.38) & Macro News (-0.26) & Regulation (-0.16) \\
		2  & Macro News (+0.62) & Regulation (-0.48) & Policy (-0.47) & Monetary (+0.27) & Labor (+0.16) \\
		3  & Labor (-0.86) & Macro News (-0.47) & Fiscal (+0.45) & Elections (+0.44) & Policy (-0.34) \\
		4  & Macro News (-0.65) & Trade (+0.46) & Elections (-0.35) & Regulation (-0.33) & Policy (-0.24) \\
		5  & Policy (-0.84) & Macro News (-0.76) & Fiscal (+0.56) & Trade (+0.55) & Labor (+0.41) \\
		6  & Labor (-0.61) & Fiscal (-0.42) & Policy (+0.35) & Macro News (-0.27) & Elections (+0.21) \\
		7  & Disease (+1.18) & Fiscal (+0.75) & Elections (+0.65) & Macro News (-0.38) & Policy (-0.28) \\
		\bottomrule
	\end{tabular}
\end{table}

The first segment (1990-1993) marks the aftermath of the early 1990s recession and is characterized by relatively low levels of market disorder, as evidenced by subdued FCIX realizations. The optimal regularization parameter remains small (\( \alpha = 0.10 \)), indicating that implied volatility was shaped by a broad and diffuse array of factors. Within this regime, sentiment related to disease and labor conditions exerts a significant negative influence, consistent with a phase of public health normalization and gradual employment recovery. In contrast, commodity-related uncertainty contributes positively, reflecting lingering concerns over inflationary pressures and global supply chain disruptions that resonated throughout the early recovery period.

The second segment (1993-1998) coincides with a period of strong macroeconomic performance, institutional credibility, and stable market expansion. As the regularization parameter increases to \( \alpha = 0.60 \), the elastic net model becomes more selective, revealing a narrower set of volatility drivers. Positive contributions from macroeconomic news suggest that steady growth reports reinforced market confidence, while the negative coefficients associated with regulation and general policy uncertainty imply that predictable governance structures alleviated risk perceptions. This transition signals a regime where implied volatility responded less to broad uncertainty and more to salient affirmations of macroeconomic strength.

The third segment (1998-2002) captures the escalation and subsequent collapse of the dot-com bubble, a period of heightened asset mispricing and elevated FCIX realizations. The sparsity increases further to \( \alpha = 0.75 \), indicating that only a few factors played dominant roles in shaping implied volatility. Negative signs on labor and macroeconomic news persist, pointing to their continued stabilizing influence. However, fiscal uncertainty and electoral dynamics emerge as significant positive contributors, suggesting that volatility perceptions became increasingly sensitive to public sector spending debates and political transitions. The prominence of electoral uncertainty during this era is particularly notable, as investor sentiment became finely attuned to anticipated policy responses to mounting financial fragility in the technology sector.

During the fourth segment (2003-2007), the market experienced a temporary moderation of volatility prior to the global financial crisis. Here, the model achieves its highest degree of sparsity (\( \alpha = 0.85 \)), signaling an intense focus on a small subset of volatility drivers. Macro news continues to play a moderating role, indicative of sustained market confidence in economic fundamentals. Nevertheless, trade, regulation, and election-related uncertainties begin to exert upward pressure on implied volatility, foreshadowing the geopolitical and institutional imbalances that would later unravel dramatically. The modest positive influence of trade-related uncertainty may be interpreted as an early reflection of concerns regarding global capital imbalances and the sustainability of international economic linkages.

The fifth segment (2007-2009) encompasses the apex of the global financial crisis, a period characterized by systemic financial instability and sharply elevated FCIX levels. The model remains sparse (\( \alpha = 0.80 \)) while achieving its highest predictive accuracy (\( R^2 = 0.917 \)). A small cadre of explanatory forces dominates this regime: macroeconomic news and policy uncertainty act as powerful suppressors of volatility, while fiscal and trade-related uncertainties emerge as potent amplifiers. Labor uncertainty, reversing its previous pattern, now assumes a positive coefficient, reflecting heightened market sensitivity to widespread employment disruptions and labor market dislocations. This shift highlights the reconfiguration of investor focus toward systemic economic fragility and sovereign response capacity during acute stress.

The sixth segment (2010-2019) represents a prolonged phase of post-crisis normalization, featuring subdued FCIX dynamics and broad-based market recovery. The regularization parameter declines to \( \alpha = 0.45 \), signifying a return to a more diversified informational environment. Labor and fiscal sentiment remain influential, both contributing negatively to implied volatility, consistent with improving employment conditions and fiscal stabilization. Nevertheless, policy and electoral uncertainties persist as positive volatility drivers, underscoring the residual influence of political dynamics even during periods of macroeconomic resilience and low realized chaos.

The final segment (2020-2023) encapsulates the unprecedented disruptions associated with the COVID-19 pandemic and its aftermath. In this high-chaos regime, the model remains relatively sparse (\( \alpha = 0.70 \)) with strong explanatory power (\( R^2 = 0.854 \)). Disease-related uncertainty emerges as the overwhelmingly dominant driver, exhibiting a large positive coefficient (+1.18), a magnitude unparalleled across all prior segments. Fiscal and electoral uncertainties also rank among the strongest volatility amplifiers, reflecting heightened sensitivity to government stimulus measures and institutional efficacy during the pandemic. Traditional macroeconomic news, while retaining a stabilizing role, proves insufficient to counterbalance the profound exogenous shocks to public health and political systems.

Comparing these segments reveals a dynamic reconfiguration of volatility drivers that tracks closely with regime shifts detected by FCIX. During tranquil periods, the model favors smaller \( \alpha \) values, capturing a broad spectrum of modest informational signals. Conversely, in periods of systemic stress, such as 2007-2009 and 2020-2023, sparsity increases sharply, highlighting the concentration of market volatility around a few decisive concerns, typically fiscal interventions, institutional stability, or exogenous health crises. These shifts coincide with surges in FCIX, reinforcing the view that heightened realized volatility is associated with a narrowing of the informational channels through which investors interpret systemic risks.

Moreover, a longitudinal analysis across all seven segments reveals a subset of EMV categories that exert persistent influence, albeit in regime-dependent ways. Macro news appears as the most consistently influential driver, featuring in all segments, though its sign and magnitude vary in accordance with broader economic conditions. Labor and policy uncertainties recur across six segments each, their impacts shifting from stabilizing forces during expansions to volatility amplifiers during contractions. Fiscal uncertainty features prominently in five segments, particularly around major financial and macroeconomic disruptions, while electoral uncertainty emerges as an influential but episodic driver, especially during periods of governance transition. Disease-related uncertainty, while concentrated within the pandemic era, underscores the market's acute sensitivity to exogenous, low-frequency shocks.

Overall, this analysis underscores that while certain categories (e.g., disease, commodities) exert episodic influence, others (macro news, policy, labor) represent structural determinants of implied volatility. Their effects, however, are highly contingent upon the prevailing market regime and economic stress levels, modulating both the dimensionality of the informational environment and the nature of volatility transmission mechanisms. This interplay between informational sparsity, FCIX dynamics, and volatility regimes offers a comprehensive perspective on how financial markets assimilate and react to evolving systemic risks.

\section{Conclusion}
\label{Sec:Conclusion}

This paper has presented a unified empirical and mathematical investigation into the structural dynamics of stock market volatility. Utilizing the FCIX, a tensor-based measure of realized volatility, we characterized the evolution of financial markets over the period January 1990 to December 2023 through a regime-switching framework grounded in the MLP distribution. Our analysis identified three distinct volatility regimes, low-chaos, intermediate-chaos, and high-chaos, each exhibiting differing levels of systemic stress, statistical dispersion, and persistence characteristics. These findings substantiate the nonstationary and regime-dependent nature of financial market behavior, while also capturing the stochastic dilation of perceived time during periods of systemic disruption.

In addition to the segmentation of realized volatility, we explored the informational forces that influence forward-looking volatility expectations. Employing sentiment-based predictors derived from the EMV tracker, we constructed elastic net regression models to forecast implied volatility, as proxied by the VIX index. The empirical results highlight that shifts in macroeconomic, fiscal, regulatory, and geopolitical uncertainty possess substantial predictive power, and that the relevance and influence of these informational categories vary systematically across distinct volatility regimes. These patterns suggest that market responses to informational shocks are highly regime-sensitive, reinforcing the necessity of modeling volatility within a segmented, regime-aware framework.

Taken together, the findings underscore that stock market volatility emerges from a dynamic interplay between endogenous structural shifts, as encoded by regime transitions in the FCIX, and exogenous informational shocks, as captured through sentiment-based predictors. The tensor-based construction of FCIX, combined with nonparametric regime detection and sparse predictive inference, provides a robust and flexible framework for diagnosing systemic fragility, identifying latent market regimes, and forecasting volatility under evolving informational conditions.

Future research directions include the development of real-time regime detection algorithms capable of operating under streaming data conditions, the integration of network-theoretic measures to capture contagion effects within financial systems, and the exploration of higher-frequency extensions of the FCIX to study intraday volatility dynamics. Moreover, further investigation into the causal pathways linking different categories of uncertainty to regime transitions could offer deeper insights into the mechanisms underlying financial market turbulence, with potential applications in systemic risk monitoring, portfolio stress testing, and macroprudential policy design.

\newpage
\bibliography{references}% common bib file

\end{document}